\documentclass[11pt,showpacs,preprintnumbers,amsmath,amssymb]{revtex4}
\usepackage{units}
\usepackage{bbold,euler,newcent}
\usepackage{graphicx}
\usepackage{dcolumn}
\usepackage{bm,bbm,mathtools,esvect}
\usepackage{slashed}

\begin{document}

\title{\LARGE\sc Quaternionic quantum mechanics \\in real Hilbert space\\\vspace{5mm}}
\author{\tt\large SERGIO GIARDINO} \vspace{1cm}
\email{sergio.giardino@ufrgs.br}
\affiliation{\vspace{3mm} Departamento de Matem\'atica Pura e Aplicada, Universidade Federal do Rio Grande do Sul (UFRGS)\\
Avenida Bento Gon\c calves 9500, 91509-900  Porto Alegre, RS, Brazil}

\begin{abstract}
\noindent A formulation of quaternionic quantum mechanics ($\mathbb{H}$QM) is presented in terms of a real Hilbert space.
Using a physically motivated scalar product, we prove the spectral theorem and obtain a novel quaternionic Fourier
series. After a brief discussion on unitary operators in this formalism, we conclude that this quantum theory is indeed consistent, and can be a valuable tool in the search for new physics.
\end{abstract}

\maketitle
\tableofcontents
\section{\sc Introduction}
Hilbert space is the standard framework used to depict wave functions in quantum mechanics. It provides 
a linear structure for physical states and an inner product for expressing physically measurable 
expectation values. On the other hand, the wave function values occur over a specific number field, and 
there 
are quantum mechanical formulations for wave functions that are evaluated over the real numbers, over the 
complex numbers and over the quaternionic numbers, but complex quantum mechanics ($\mathbb{C}$QM) is the 
standard and most important quantum theory. Supposing a correspondence between the elements of the 
Hilbert space and of its inner product, a complex wave function requires a complex scalar product as 
well as a complex Hilbert space, so that an identical rule would be valid for any other number field in 
which wave functions take their values. In this article, we provide an example of a quantum theory where 
wave functions and expectation values belong to different number fields, which is physically interesting 
and mathematically consistent.

The $\mathbb{C}$QM have been developed, with John von Neumann and Garrett Birkhoff being the first to 
figure out the existence of real and quaternionic formulations for QM \cite{Birkhoff:2017kpl}.   
Ernst Stueckelberg achieved the development of real quantum mechanics ($\mathbb{R}$QM)
\cite{Stueckelberg:1960rsi,Stueckelberg:1961rsi,Stueckelberg:1961psg,Stueckelberg:1962fra}, and this construct
is considered equivalent to $\mathbb{C}$QM. However, Stueckelberg's formulation is involved; it demands
an anti-unitary operator $J$ that replaces the imaginary unit $i$ in the real Schr\"odinger equation, and this 
anti-unitary operator is specific for each anti-commuting pair of operators \cite{Stueckelberg:1960rsi}. In 
spite of this difficulty, $\mathbb{R}$QM is still an object of research, and plays a role within quantum information 
\cite{Mosca:2009sqs,Borish:2013rqm,Wootters:2014ret,Wootters:2012esh} and mathematical physics 
\cite{Oppio:2016pbf}, thus indicating that alternatives to $\mathbb{C}$QM may be necessary to understand new physical phenomena.

Another possibility for testing the structure of quantum mechanics is to replace the complex numbers with 
the quaternionic numbers. We briefly remember that the quaternions ($\mathbb{H}$) are hyper-complex 
numbers with three imaginary units. If we use $q\in\mathbb{H}$, then we obtain
\begin{equation}\label{i1}
 q=x_0 + x_1 i + x_2 j + x_3 k, \qquad\mbox{where}\qquad x_0,\,x_1,\,x_2,\,x_3\in\mathbb{R}\qquad\mbox{and}\qquad i^2=j^2=k^2=-1.
\end{equation}
The imaginary units $i,\,j$ and $k$ are anti-commutative (by way of example, $ij=-ji$), so that 
quaternions are non-commutative hyper-complex numbers. A quaternionic number may also be written in 
symplectic notation by using the complex components $\,z_0,\,z_1\in\mathbb{C},\,$, so that 
$\,q=z_0+z_1j.\,$  Additional information on quaternions and quaternionic physics may be found in 
\cite{Rocha:2013qtt}. Accordingly, if the quaternions generalize the complex, quaternionic wave 
functions may generalize complex wave functions in quantum mechanics, providing additional degrees of
freedom to the theory. This is a physical motivation to study such a quaternionic theory. 

In practice, a quaternionic quantum theory was attained by using anti-hermitian operators, a theory 
that is mainly depicted in a book by Stephen Adler \cite{Adler:1995qqm}. We remember that an 
anti-hermitian operator $\mathcal{A}$ and its adjoint operator $\mathcal{A}^\dagger$ satisfy 
$\mathcal{A}^\dagger=-\mathcal{A}$. The anti-hermitian formulation of $\mathbb{H}$QM displays several 
drawbacks, such as a badly defined classical limit; exact solutions are scarce, complicated and 
difficult to interpret physically 
\cite{Davies:1989zza,Davies:1992oqq,Ducati:2001qo,Nishi:2002qd,DeLeo:2005bs,Madureira:2006qps,Ducati:2007wp,Davies:1990pm,DeLeo:2013xfa,DeLeo:2015hza,Giardino:2015iia,Sobhani:2016qdp,Procopio:2016qqq}. 
In this article, we discuss the mathematical basis of a $\mathbb{H}$QM without anti-hermitian operators, 
and we believe that a simpler theory emerges in this case. In order to introduce the basic ideas of such 
a theory, let us entertain the quaternionic Schr\"odinger equation 
\begin{equation}\label{i2}
 \hslash\frac{\partial\Psi}{\partial t}\,i\,=\mathcal{H}\,\Psi,
\end{equation}
where $\mathcal{H}$  is the Hamiltonian operator. We emphasize that the imaginary unit $\,i\,$ stands 
at the right-hand side of $\,\partial_t\Psi,\,$ because of $\,i\Psi\neq\Psi i$. In symplectic notation, 
the quaternionic wave function reads
\begin{equation}\label{i3}
 \Psi=\Psi_0+\Psi_1\,j,
\end{equation}
where $\Psi_0$ and $\Psi_1$ are complex functions, the quaternionic Hamiltonian is
\begin{equation}\label{i4}
\mathcal H=-\frac{\hslash^2}{2m}\Big(\bm\nabla- \bm{\mathcal A}\Big)^2+U,\qquad
\mbox{where}\qquad\bm{\mathcal A}=\bm\alpha i+\bm\beta j,\qquad U=V +W\,j,
\end{equation}
where $\bm\alpha$ is a real vector function, $\bm\beta$ is a complex vector function and $V$ and $W$ 
are complex scalar functions. Since $\mathcal{H}^\dagger\neq\mathcal H$, the Hamiltonian operator is neither
hermitian nor anti-hermitian, and consequently the usual frameworks of hermitian $\mathbb{C}$QM and anti-hermitian
$\mathbb{H}$QM do not seem applicable to (\ref{i2}-\ref{i4}). We may hypothesize non-hermitian $\mathbb{C}$QM 
\cite{Bender:2007nj,Moiseyev:2011nhq} as an appropriate way for understanding $\mathbb{H}$QM through (\ref{i2}-\ref{i4});
however, a reliable approach to this hypothesis requires several consistency conditions to be met. As a first consistency test,
let us entertain the conservation of the probability, depicted in the continuity equation  \cite{Giardino:2016abe,Giardino:2017pns}, 
\begin{equation}\label{i5}
\frac{\partial \rho}{\partial t}+ \bm{\nabla\cdot J}=g,
\end{equation}
with $\rho=\Psi\overline\Psi^{\,}$ as the probability density, $\bm J$ as the probability current and $g$ as a probability source, namely
\begin{equation}\label{i6}
g=\frac{1}{\hslash}\Big(\Psi i\,\overline\Psi^{\,}\bar U-U\Psi i\overline\Psi^{\,} \Big),
\;\;\;\;
\bm J=\frac{1}{2m}\Big[(\bm\Pi\Psi)\overline\Psi^{\,}+\Psi\big(\,\overline{\bm\Pi\Psi}\,\big) \Big]\;\;\;\;\mbox{and}\qquad
\bm\Pi\Psi=-\hslash\big(\bm\nabla-\bm{\mathcal A}\big)\Psi i.
\end{equation}
We point out that $\bar q=\bar z_0-z_1j$ is the quaternionic conjugate of $q=z_0+z_1j$,  and that 
$\bm\Pi$ is the generalized momentum operator. Equation (\ref{i5}) preserves the probability density when 
$\,g=0\,$, and this occurs for real $U$; furthermore complex potentials, where $W=0$,  recover a 
continuity equation for non-hermitian $\mathbb{C}$QM, irrespective of $\bm{\mathcal A}$. These simple 
facts show that the model is statistically consistent and is physically related to  well-established 
results of $\mathbb{C}$QM. In other words, this model is a simple extension of the complex quantum theory, 
and our aim is to examine its mathematical consistency.

The conservation of probability may be considered a previous consistency check of the classical limit, 
since probability is strongly related to classical physics. In order to ascertain the classical limit of 
$\mathbb{H}$QM, the definition of the  formula of an expectation value is needed. Anti-hermitian 
$\mathbb{H}$QM takes the expectation value of the complex theory for granted, but this is only a possible 
choice. We alternatively obtain the expectation value from the linear momentum expectation value 
\cite{Giardino:2016abe}, that is related to the probability current through 
$\langle \bm \Pi\rangle=\langle \bm J\rangle/m$. Accordingly, we define the expectation value for an 
arbitrary quaternionic operator $\mathcal{O}$  as
\begin{equation}\label{i7}
\langle\mathcal O\rangle= \frac{1}{2}\int dx^3\Big[\big(\mathcal{O}\Psi\big)\overline\Psi^{\,} +\Psi\big(\overline{\mathcal{O}\Psi}\big) \Big].
\end{equation}
This expectation value is real irrespective of $\mathcal O$, a quality that is physically desired. 
Furthermore, (\ref{i7}) recovers the $\mathbb{C}$QM expectation value for complex wave functions and 
hermitian operators. The above definition also allows well defined classical limits 
\cite{Giardino:2017pns}, a clear advantage over anti-hermitian $\mathbb{H}$QM. Nevertheless, there is a 
primary dissimilarity to the usual definition of expectation values in $\mathbb{C}$QM, since (\ref{i7}) 
imposes a real inner product on the Hilbert space, and consequently the Hilbert space becomes a real 
linear space, openly contradicting the picture of quantum theories where the inner product and the wave 
function are evaluated over identical fields. Presumably, there are profound consequences for this fact in 
$\mathbb{H}$QM, now seen as a theory for quaternionic wave functions in a real Hilbert space, and we will 
examine them in the next section.
\section{\sc Real Hilbert space\label{R}}
From (\ref{i7}), we propose that the inner product between two quaternionic functions $\Phi$ and 
$\Psi$ is 
\begin{equation}\label{r1}
\big\langle\mathcal\Phi,\,\Psi \big\rangle= \frac{1}{2}\int dx^3\Big(\Phi\overline{\Psi}^{\,} +\Psi\overline{\Phi}^{\,} \Big),
\end{equation}
which immediately fulfills $\,\langle \Psi,\,\Psi\rangle>0.\,$ The inner product (\ref{r1}) also conforms
\begin{align}
\label{r2}& \big\langle\Phi,\,\Psi \big\rangle =\big\langle\Psi,\,\Phi \big\rangle,\\
\label{r3}& \big\langle \alpha\,\Phi+\beta\,\Xi,\,\Psi\big\rangle=\alpha\,\big\langle\Phi,\,\Psi\big\rangle+\beta\,\big\langle\Xi,\,\Psi \big\rangle,\\
\label{r4}& \big\langle\Psi,\,\Psi \big\rangle=\big\|\Psi\big\|^2;
\end{align}
for $\,\alpha,\,\beta\in\mathbb{R},\,$ and $\|\Psi\|$ the norm of $\Psi$; thus we have a real vector 
space. If the vector space is complex, then 
$\big\langle\Phi,\,\Psi \big\rangle =\overline{\big\langle\Psi,\,\Phi \big\rangle}$ replaces (\ref{r2}) 
and conversely the inner product of the complex linear space is sesquilinear, so that 
$\big\langle\Phi,\,\alpha\,\Psi \big\rangle=\big\langle\Phi,\,\Psi \big\rangle\bar\alpha$ holds. In order 
to obtain a real Hilbert space, let us consider several necessary properties. First of all, the Schwarz 
inequality
\begin{equation}\label{r5}
 \big|\big\langle\Phi,\,\Psi\big\rangle\big|\leq\big\|\Psi\big\|\,\big\|\Phi\big\|.
\end{equation}
Secondly, the joint continuity of the inner product, where
\begin{equation}\label{r6}
\mbox{if}\qquad \Psi_n\to\Psi\qquad\mbox{and}\qquad\Phi_n\to\Phi\qquad\Rightarrow\qquad\big\langle\Phi_n,\,\Psi_n\big\rangle\to
\big\langle\Phi,\,\Psi\big\rangle.
\end{equation}
Another important condition is the parallelogram law
\begin{equation}\label{r7}
 \big\|\Phi+\Psi\big\|^2+ \big\|\Phi-\Psi\big\|^2=2 \big\|\Phi\big\|^2+2 \big\|\Psi\big\|^2,
\end{equation}
and finally, the orthogonality
\begin{equation}\label{r8}
\big\langle\Phi,\,\Psi\big\rangle=0\qquad\Rightarrow\qquad\Phi\perp\Psi.
\end{equation}
The proofs of (\ref{r5}-\ref{r8}) are straightforward and may be found  in many textbooks, we quote 
\cite{Simmons:2004ita} as an example. We can accordingly define the complete orthonormal basis 
$\,\{ \Lambda_a\}\,$ into the real Hilbert space $H$, so that the following
statements are all equivalent one to another
\begin{align}
 &\{\Lambda_i\}\qquad\mbox{is complete};\\
& \Omega\perp\{ \Lambda_a\}\qquad\Rightarrow\qquad \Omega=0\,;\\
&\forall\, \Omega\,\in\, H, \qquad\Rightarrow\qquad \Omega=\sum_a \big\langle\Omega,\,\Lambda_a \big\rangle\Lambda_a\,;\\
&\forall\, \Omega\,\in\, H, \qquad\Rightarrow\qquad \|\Omega\|^2=\sum_a \big|\big\langle\Omega,\,\Lambda_a \big\rangle\big|^2,
\end{align}
and we adduce that the basis may be either finite or infinite. A real Hilbert space of quaternionic 
functions fulfills  these necessary requirements. However, this conclusion appears to be a mathematical 
curiosity of academic interest only. In order to apply this formalism to quantum mechanics, we need a 
concrete representation of a real Hilbert space of quaternionic functions. In the next section we provide
such a construct.
\section{\sc Quaternionic Fourier expansion}
As a prototype, let us consider the $\mathcal{L}^2[0,\,2\pi]$ space of square integrable complex functions
defined on $[0,\,2\pi]$. The inner product for this space is
\begin{equation}\label{q1}
 (f,\,g)=\intop_0^{2\pi}f(x)\bar g(x),\qquad\forall\, f,\,g\in\mathcal{L}^2[0,\,2\pi],
\end{equation}
and the square integrability means that $\;\|f(x)\|^2<\infty.\;$ A base for this space deploys complex 
exponentials and satisfies the orthogonality condition
\begin{equation}\label{q2}
 \big(e_m,\,e_n\big)=\,\delta_{mn},\qquad\mbox{where}\qquad e_n=\frac{1}{\sqrt{2\pi}}e^{inx}\qquad\mbox{and}\qquad m,n\in\mathbb{Z}.
\end{equation}
The completeness of the base means that each $f\in\mathcal{L}^2[0,\,2\pi]$ admits a Fourier expansion
\begin{equation}\label{q3}
f=\sum_{n=-\infty}^{\infty}\big(f,\,e_n\big)\,e_n.
\end{equation}
Our aim is to obtain a quaternionic analogue of the above complex case. Using (\ref{r1}), we define 
the inner product as
\begin{equation}\label{q4}
\big\langle\Phi,\,\Psi \big\rangle= \frac{1}{2}\int_0^{2\pi} dx^3\Big(\Phi\overline{\Psi}^{\,} +\Psi\overline{\Phi}^{\,} \Big), \qquad\mbox{for}\qquad
\|\Phi\|^2<\infty\qquad\mbox{and}\qquad\|\Psi\|^2<\infty.
\end{equation}
In order to obtain the Fourier expansion for quaternionic functions, we propose an orthogonal basis made 
of unitary quaternions, where
\begin{equation}\label{q5}
 \Lambda=\cos\theta e^{i\phi}+\sin\theta e^{i\xi}\,j,\qquad\mbox{so that}\qquad 
\Lambda\bar\Lambda=1.
\end{equation}
The identity
\begin{equation}\label{q6}
 \mathfrak{Re}\Big[\Lambda\,\Lambda'\Big]=\cos\theta\cos\theta'\cos(\phi-\phi')+\sin\theta\sin\theta'\cos\big(\xi-\xi'),
\end{equation}
with $\Lambda'=\Lambda(\theta',\,\phi'\,\,\xi')$ permits us to build quaternionic basis sets that are 
analogous to the usual basis of complex Fourier series (\ref{q3}). Let us entertain the quaternionic 
Fourier series
\begin{equation}\label{q7}
F(x)=\sum_{n=-\infty}^{\infty} a_n \Lambda_n,
\end{equation}
to which we propose the quaternion unitary basis
\begin{equation}\label{q8}
\Lambda_n=\cos nx\, e^{i\phi_0}+\sin nx\, e^{i\xi_0}j,\qquad\mbox{with}\qquad n\in\mathbb{Z},\qquad x\in[0,\,2\pi],
\end{equation}
and $\,\phi_0,\,\xi_0\,$ constants. We cannot eliminate the negative indexes because $\Lambda_n$
does not have a definite parity, like when sine series or cosine series are considered in isolation. 
The coefficients of expansion $a_n$ are real because we adopt a real inner product. Thus, using 
(\ref{q4}) and (\ref{q6}), we obtain
\begin{equation}\label{q9}
\big\langle\Lambda_n,\,\Lambda_{n'} \big\rangle=2\pi\,\delta_{nn'}.
\end{equation}
This result permits us to calculate the $a_n$ coefficients
\begin{equation}\label{q10}
a_n=\big\langle F(x),\,\Lambda_n\big\rangle=\frac{1}{\sqrt{2\pi}}\int_0^{2\pi}\mathfrak{Re}\big[\,F(x) \Lambda_n\big],
\end{equation}
which of course defines a quaternionic Fourier transformation. There are different quaternionic bases, 
for example
\begin{equation}\label{q11}
\Lambda_n=\cos\theta_0\, e^{inx}+\sin\theta_0\, e^{-inx}j,\qquad\mbox{with}\qquad n\in\mathbb{Z},\qquad x\in[0,\,2\pi],
\end{equation}
with a constant $\,\theta_0.\,$ The (\ref{q11}) basis set satisfies the  orthogonality condition (\ref{q9})
as well. Surprisingly, these bases preserve their orthogonality when $\theta_0,\,\phi_0$ and $\xi_0$ are 
turned into arbitrary functions. On the other hand, we may imagine additional indexes for the basis sets. 
By way of example, either
\begin{equation}\label{q12}
\Lambda_{mn}=\cos\theta_0\, e^{imx}+\sin\theta_0\, e^{inx}j,\qquad\mbox{or}\qquad
\Lambda_{\ell mn}=\cos\ell x\, e^{imx}+\sin\ell x\, e^{inx}j
\end{equation}
may be resolved to find basis elements. In the orthogonality conditions we obtain additional Kronecker 
deltas, and thus (\ref{q10}) turns into an infinite linear system. Solutions to such kind of systems are 
known in a variety of cases \cite{Winkler:1979hse,Gohberg:tdl,Simon:2010tit}, and the important thing here 
is only to point out that there are solutions to the problem of determining a basis for the space. The 
investigation of such multi-indexed bases is, of course, a very interesting direction for future research.

To the extent of our knowledge, this is the first proposal for a quaternionic Fourier series, and the 
consequent quaternionic Fourier transform  is also novel. Quaternionic series are a subtle and difficult 
matter, and the holomorphic power quaternionic series only allows for linear series
\cite{Sudbery:1979qta,Deavours:1973qtc}. There are other proposals where more sophisticated series are 
allowed, for example the so called regular quaternionic series \cite{GSS}, but in fact the quaternionic 
power series are not trivial. The Fourier quaternionic series (\ref{q7}) is surprisingly simple, and this 
simplicity turns  it into an interesting and suitable device for applications without the need of highly 
sophisticated mathematical tools. We hope that exact solutions in $\mathbb{H}$QM may be expressed using 
these kinds of series, and this is of course a very interesting question for future investigations.

\section{\sc The spectral theorem}

The spectral theorem follows straightforwardly from the features of the Hilbert space: it is complete, has
a well-defined inner product, and is normed. As a consequence, the conjugate space is accordingly 
well-defined, and there is a unique vector $y$ in the Hilbert space corresponding to an arbitrary 
functional $f(x)$, so that $f(x)=\langle x,\,y\rangle$. Adjoint operators and projections are also 
well-defined. There are differences, however. Considering a quaternionic normal operator $N$, which 
commutes with its adjoint $N^\dagger$, we can write a symplectic decomposition, so that
\begin{equation}\label{s1}
N=N_0+N_1 j\qquad\mbox{and}\qquad N^\dagger=N_0^\dagger-N_1j,
\end{equation}
where we assumed that $N_k j=jN_k^\dagger$, for $k=0,\,1$. The decomposition (\ref{s1}) is motivated by 
analogy with the complex case, where normal operators have a structure that is analogous to the complex 
numbers. Imposing the normal condition
\begin{equation}\label{s2}
\big[N,\,N^\dagger\big]=0,
\end{equation}
we obtain two conditions
\begin{equation}\label{s3}
\big[N_0,\,N_0^\dagger\big]=0,\qquad\mbox{and}\qquad \big[N_0+N_0^\dagger,\,N_1\big]=0.
\end{equation}
Thus, $N_0$ is a normal operator of a complex Hilbert space, and therefore $N$ has a well behaved
complex limit within $N_1\to 0$, as expected. On the other hand the pure quaternionic component $N_1$ is 
not necessarily a normal operator, but it must commute with the ``real component'' of $N_0$. Thus, we 
have quaternionic normal operators, whose existence is important because self-adjoint operators 
form a subset of them. Further important operators are the projections and the unitary operators. 
The existence of both of these is a current property of real vector spaces, and the spectral theorem 
follows accordingly. The proof of the theorem is straightforward, and we only outline it here. Let us 
consider an operator $T$ and its eigen-value equation
\begin{equation}\label{s4}
T\bm x=\lambda\bm x,
\end{equation}
where $\bm x\in H$ is the eigen-vector and $\lambda$ is a scalar constant. If the operator $T$ admits a finite number of eigen-vectors and 
corresponding eigen-values, we may generally write 
\begin{equation}\label{s5}
\bm x=\sum_{k=1}^n \bm x_k\qquad\mbox{so that}\qquad T\bm x=\sum_{k=1}^n \lambda_k\bm x_k
\end{equation}
where $x_k$ are the orthogonal components of $x$. Consequently the spectral resolution of the operator is obtained, namely
\begin{equation}\label{s6}
T=\sum_{k=1}^n\lambda_k P_k.
\end{equation}
$P_k$ are the projection operators, such that
\begin{equation}\label{s7}
P_k x=x_k\qquad\mbox{and}\qquad P_k x_\ell=\delta_{k\ell}x_\ell.
\end{equation}
All of the above results for real Hilbert spaces are straightforward and are depicted in many textbooks, 
we mention \cite{Simmons:2004ita} by way of example. 

The most interesting matter in this section is the interpretation of the spectral theorem. In 
$\mathbb{C}$QM, if an operator has an eigen-function with a complex eigen-value, this eigen-function still has an expression in the basis 
of the Hilbert space, although such an operator is not a physical observable. Although not observable, eigen-functions of complex eigen-values 
may have physical interpretations either in terms of non-hermitian QM or in terms of symmetries. 

In $\mathbb{H}$QM  the situation is quite different. Every operator is observable because there is a real 
expectation value. On the other hand, an eigen-function of an operator with a quaternionic eigen-value 
does not belong to the Hilbert space, and does not have an expression in terms of its basis using 
quaternionic eigenvalues. As only real eigen-values are admitted, we cannot consider (\ref{s4}) an 
eigen-value equation for quaternionic $\lambda$, and indeed me must consider $\lambda$ as an operator. 
Otherwise, the situation is physically and mathematically meaningless. We expect that this kind of 
situation is related to the symmetries of the wave function, however its features deserve an independent 
investigation in order to be properly ascertained. The question about quaternionic symmetries is deeply 
related to unitary operators, which we briefly consider in the next section.

\section{\sc Time evolution}
Unitary operators are fundamental for quantum mechanics, inasmuch as symmetry operators are associated 
with them. A unitary operator $U$ obeys the fundamental properties
\begin{equation}\label{u1}
U U^\dagger=\mathbb{1},\qquad\qquad \langle Ux,\,Uy\rangle,\qquad\qquad \|Ux,\,Ux\|=\|x\|;
\end{equation}
where $\mathbb{1}$ is an identity operator, and $x,\,y$ are arbitrary vectors of the Hilbert space. From 
the study of complex Hilbert spaces, eigen-values of unitary operators are unitary complex numbers, and 
thus unitary operators do not represent physical observables in $\mathbb{C}$QM, a well known fact. 
Consequently, unitary operators are associated with the internal properties of the wave-functions known as 
symmetries, that unravels peculiar properties of quantum mechanics. 

The first application of unitary operators in quantum mechanics involves time evolution, that may of 
course be understood as a symmetry. Supposing that a unitary operator determines the time evolution of a 
quantum wave function, the time evolved wave function will be
\begin{equation}\label{u2}
\Psi(t+\delta t)=U(t,\delta t)\Psi(t).
\end{equation}
Using the series expansions
\begin{equation}\label{u3}
U(t,\,\delta t)=1-\delta t \mathcal{A}\qquad\mbox{and}\qquad \Psi(t+\delta t)=\Psi(t)+\delta t\frac{\partial\Psi(t)}{\partial t},
\end{equation}
we get
\begin{equation}\label{u4}
\frac{\partial\Psi(t)}{\partial t}=-\mathcal{A}\Psi.
\end{equation}
In $\mathbb{C}$QM, we impose $\mathcal{A}=i\mathcal{H}/\hslash$ and recover the Schr\"odinger equation, 
while in anti-hermitian $\mathbb{H}$QM it is supposed that $\mathcal{A}$ is an anti-hermitian Hamiltonian,
from which the whole theory is developed \cite{Adler:1995qqm}. Here, we do not follow this anti-hermitian 
interpretation for $\mathbb{H}$QM, and now we explain our reasons for that. 

The formal calculation that indicates $\mathcal{A}$ as an anti-hermitian operator also holds for $\mathbb{C}$QM. Following the time evolution pattern of $\mathbb{C}$QM we introduce a unitary operator $U(t,\,t_0)$ that satisfies the time-dependent Schr\"odinger equation (\ref{i2}), namely
\begin{equation}\label{u5}
U(t,\,t_0) =\left[\mathbb{1}+\frac{1}{\hslash}\int_{t_0}^t dt'\,\mathcal{H}(t')\,U(t',\,t_0)\right](-i).
\end{equation}
Using the notation $(a|b)\Psi=a\Psi b$, we iteratively obtain
\begin{equation}\label{u6}
U(t,\,t_0) =\mathbb{1}+\sum_{n=1}^\infty\int_{t_0}^{t} dt_1\,\left(\frac{\mathcal{H}(t_1)}{\hslash}\Big|-i\right)
\int_{t_0}^{t_1} dt_2\,\left(\frac{\mathcal{H}(t_2)}{\hslash}\Big|-i\right)\;\dots\;
\int_{t_0}^{t_{n-1}} dt_n\,\left(\frac{\mathcal{H}(t_n)}{\hslash}\Big|-i\right).
\end{equation}
We define the time-ordering operator $P$, sum the permutations of each term of (\ref{u6}) and thus obtain
\begin{equation}\label{u7}
U(t,\,t_0)=P \exp \left[\,\int_{t_0}^t\left(\frac{\mathcal{H}(t')}{\hslash}\Big|-i\right)dt'\right].
\end{equation}
The above calculation, to obtain the exponential rule of unitary time evolution, is available in many textbooks of quantum mechanics, and thus we omit the details. More importantly, it seems that (\ref{u7}) could reconcile $\mathbb{H}$QM over real Hilbert spaces to anti-hermitian $\mathbb{H}$QM. However,  
considering a quaternionic $\Psi(t)=U(t,\,t_0)\Psi(t_0)$, (\ref{i2}) furnishes
\begin{equation}\label{u8}
\hslash\frac{\partial U(t,\,t_0)}{\partial t}\Psi(t_0)i=\mathcal{H} \,U(t,\,t_0)\Psi(t_0).
\end{equation}
A quaternionic $\Psi(t_0)$ does not factors out (\ref{u8}), meaning that (\ref{u5}-\ref{u7}) holds for 
the complex $\Psi(t_0)$ only. This result indicates that (\ref{u2}-\ref{u4}) are too naive and do not hold 
for arbitrary unitary quaternionic time evolution operators.

We obtain a partial reconciliation between the quaternionic quantum theories for the particular case of 
a complex $\Psi(t_0)$. A comprehensive conciliation seems impossible because of the non-commutativity of 
unitary quaternions (\ref{q5}), where 
\begin{equation}\label{u9}
\Lambda(\theta,\,\phi,\,\xi)\,\Lambda(\theta',\,\phi',\,\xi')\;\neq\;\Lambda(\theta+\theta',\,\phi+\phi',\,\xi+\xi').
\end{equation}
This simple fact indicates that $\mathbb{H}$QM permits non-abelian unitary time evolutions, something that has never been considered
in $\mathbb{C}$QM or in anti-unitary $\mathbb{H}$QM, and comprises a fascinating direction for future research.

\section{\sc Conclusion}

In this article we studied the basic equations that found a quantum mechanical theory where the wave 
functions are quaternionic and the Hilbert space is real. We conclude that the mathematical framework 
established in the complex theory  is similar to the quaternionic theory, but important differences exist. 
These differences come from the definition of the expectation value, which demands a real Hilbert space
in $\mathbb{H}$QM in contrast to the complex Hilbert space and complex inner product of $\mathbb{C}$QM. 
The real inner product enables us to obtain a consistent classical limit \cite{Giardino:2017pns}, 
something that is impossible for anti-hermitian $\mathbb{H}$QM where the Hilbert space is quaternionic.

After the establishment of $\mathbb{H}$QM in this real Hilbert space, this article adduces the proof of 
the spectral theorem in the real formulation, and additionally provides the realization of such a real 
Hilbert space in terms of the quaternionic Fourier series. This result means that quaternionic wave 
functions are not just formal possibilities, but that the elements for constructing these kinds of 
quantum solutions are already available. The description of unitary operators is a further topic where the
real Hilbert space $\mathbb{H}$QM and the anti-hermitian theory are profoundly different. This is mainly 
because the quaternionic theory in the real Hilbert space admits a time evolution described by a 
non-abelian unitary operator, while the anti-hermitian theory justifies an unitary time evolution from 
(\ref{u2}-\ref{u4}). We contended that such arguments are not valid in general, and quaternionic unitary 
operators must be, of course, non-abelian. A time evolution in terms of non-commutative unitary operators 
has never been described in $\mathbb{C}$QM, mainly because this kind of evolution is not grounded by 
Stone's theorem. However, it is a natural possibility in quaternionic theory, and consequently it 
comprises a fundamental direction for future research.

In conclusion, the results indicate that $\mathbb{H}$QM in a real Hilbert space is a consistent theory, 
constituting an environment where new solutions may be researched without the fear of a fundamental 
inconsistency. This is the most important conclusion of this article. There are various directions for 
future research, and almost every issue of $\mathbb{C}$QM may be studied within this 
quaternionic formalism. Particular solutions of the quaternionic Schr\"odinger equation, in terms of the
quaternionic Fourier series, are an interesting  subject, mostly because they improve the 
physical understanding throughout examples of applications. The investigation of quaternion quantum 
symmetries is another matter of interest. In summary, real $\mathbb{H}$QM is a novel and consistent 
quantum theory. The next stage is presumably to investigate whether it can describe something 
physically interesting, and hopefully also describe phenomena that cannot be described in terms of 
$\mathbb{C}$QM. If true, this last possibility would turn $\mathbb{H}$QM into a formalism for 
new physics, with this very last possibility being the main motivation for further research into
$\mathbb{H}$QM.


%
%
%
%

\bibliographystyle{unsrt} 
\bibliography{bib_hilbert}

\end{document}